\documentclass[pdflatex,prl,twocolumn,showpacs,superscriptaddress]{revtex4-1}
\usepackage{bm,amsmath,amssymb}
\usepackage{graphicx}
\usepackage{subfigure}
\usepackage{color}
\usepackage{soul}
\usepackage{hyperref}
\usepackage{multirow}
\usepackage{lineno}
\setulcolor{blue}
\setstcolor{red}
\sethlcolor{yellow}

\newcommand \eg {{\sl e.g.}}
\newcommand \ie {{\sl i.e.}}

\newcommand \fig[1] {Fig.\ \ref{#1}}

\newcommand \eq[1] {Eq.\ (\ref{#1})}
\newcommand \beq {\begin{equation}}
\newcommand \eeq {\end{equation}}
\newcommand \beqa {\begin{eqnarray}}
\newcommand \eeqa {\end{eqnarray}}

\newcommand \nn {\nonumber}
\newcommand \hmu {\hat{\mu}}
\begin{document}

\title{Strangeness at high temperatures: from hadrons to quarks}

\author{A. Bazavov}
\affiliation{Physics Department, Brookhaven National Laboratory, Upton, NY 11973, USA}
\author{H.-T. Ding}
\affiliation{Physics Department, Brookhaven National Laboratory, Upton, NY 11973, USA}
\affiliation{Physics Department, Columbia University, New York, NY 10027, USA}
\author{P. Hegde}
\affiliation{High Energy Physics Lab, Department of Physics R518, National Taiwan
University, Taipei 10617, Taiwan}
\author{O. Kaczmarek}
\affiliation{Fakult\"at f\"ur Physik, Universit\"at Bielefeld, D-33615 Bielefeld,
Germany}
\author{F. Karsch}
\affiliation{Physics Department, Brookhaven National Laboratory, Upton, NY 11973, USA}
\affiliation{Fakult\"at f\"ur Physik, Universit\"at Bielefeld, D-33615 Bielefeld,
Germany}
\author{E. Laermann}
\affiliation{Fakult\"at f\"ur Physik, Universit\"at Bielefeld, D-33615 Bielefeld,
Germany}
\author{Y. Maezawa}
\affiliation{Physics Department, Brookhaven National Laboratory, Upton, NY 11973, USA}
\author{Swagato Mukherjee}
\affiliation{Physics Department, Brookhaven National Laboratory, Upton, NY 11973, USA}
\author{H. Ohno}
\affiliation{Fakult\"at f\"ur Physik, Universit\"at Bielefeld, D-33615 Bielefeld,
Germany}
\author{P. Petreczky}
\affiliation{Physics Department, Brookhaven National Laboratory, Upton, NY 11973, USA}
\author{C. Schmidt}
\affiliation{Fakult\"at f\"ur Physik, Universit\"at Bielefeld, D-33615 Bielefeld,
Germany}
\author{S. Sharma}
\affiliation{Fakult\"at f\"ur Physik, Universit\"at Bielefeld, D-33615 Bielefeld,
Germany}
\author{W. Soeldner}
\affiliation{Institut f\"ur Theoretische Physik, Universit\"at Regensburg, D-93040
Regensburg, Germany}
\author{M. Wagner}
\affiliation{Fakult\"at f\"ur Physik, Universit\"at Bielefeld, D-33615 Bielefeld,
Germany}

\begin{abstract}

Appropriate combinations of up to fourth order cumulants of net strangeness
fluctuations and their correlations with net baryon number and electric charge
fluctuations, obtained from lattice QCD calculations, have been used to probe the
strangeness carrying degrees of freedom at high temperatures. For temperatures up to
the chiral crossover separate contributions of strange mesons and baryons can be well
described by an uncorrelated gas of hadrons. Such a description breaks down in the
chiral crossover region, suggesting that the deconfinement of strangeness takes place
at the chiral crossover. On the other hand, the strangeness carrying degrees of
freedom inside the quark gluon plasma can be described by a weakly interacting gas of
quarks only for temperatures larger than twice the chiral crossover temperature. In
the intermediate temperature window these observables show considerably richer
structures, indicative of the strongly interacting nature of the quark gluon plasma.

\end{abstract}


\maketitle

{\emph{Introduction.---} Strangeness has played a crucial role \cite{strange-rev} in
the experimental and theoretical investigations of the deconfined  phase, namely the
Quark Gluon Plasma (QGP) phase, of Quantum Chromodynamics (QCD) at high temperatures.
Experimental results from the Relativistic Heavy Ion Collider and the Large Hadron
Collider suggest that the QGP has been created during the highly energetic collisions
of heavy nuclei \cite{hic-rev}. Experimental results showing enhanced production and
large collective flow of strange hadrons \cite{strange-expt} strongly indicate that
deconfined strange quarks existed inside the QGP, despite the absence of real strange
quarks within the initially colliding nuclei.  However, theoretical understanding of
the deconfinement of strangeness remains unclear. A-priori it is not unreasonable to
expect that the heavier strange quark may not be largely influenced by the chiral
symmetry of QCD and the deconfinement of the strange quarks may not take place at the
chiral crossover temperature ($T_c$).  Based on the observations that, compared to
the light up and down quarks, the net strange quark number fluctuations
\cite{WHRG,hotQCDHRG} show a much smoother behavior across the chiral crossover
region, it has been suggested \cite{WB-Tc} that the deconfinement crossover for the
strange quarks may take place at a temperature larger than $T_c$. Consequently
strange hadronic bound states may exist inside the QGP for temperatures $T\gtrsim
T_c$ \cite{Ratti}. 

Moreover, the nature of the deconfined QGP for moderately high temperatures also
remains elusive. An intriguing open question is whether in this temperature regime
the QGP is a strongly coupled medium lacking a quasi-particle description
\cite{hic-rev} or consists of other degrees of freedom such as colored bound states
\cite{Shuryak} or massive colored quasi-particles \cite{quasipart1}.  Knowledge
regarding the behavior of strangeness carrying Degrees of Freedom (sDoF) in the QGP
is essential to answer this question.

It is well known \cite{Koch,Ejiri} that the quantum numbers, such as the baryon
number ($B$), electric charge ($Q$) and strangeness ($S$), can be probed using the
fluctuations and correlations of these quantities . We construct observables from
combinations of up to fourth order cumulants of net strangeness fluctuations and
their correlations with net baryon number and electric charge fluctuations that probe
the sDoF in different temperature regimes. We calculate these observables using
state-of-the-art Lattice QCD (LQCD) simulations and compare our results with the
hadron gas description at lower temperatures and with the weakly interacting quark
gas description at higher temperatures.


\emph{Strangeness in a gas of uncorrelated hadrons.---} For an uncorrelated gas of
hadrons, \eg\ the Hadron Resonance Gas (HRG) model \cite{pbm}, the dimensionless
partial pressure, $P_{S}\equiv(p-p_{S=0})/T^4$, of all the strange hadrons is given
by 
\beqa
P^{HRG}_S(\hmu_B,\hmu_S) &=& P^{HRG}_{|S|=1,M} \cosh(\hmu_S) 
\nn \\
&+& P^{HRG}_{|S|=1,B} \cosh(\hmu_B-\hmu_S)
\nn \\
&+& P^{HRG}_{|S|=2,B} \cosh(\hmu_B-2\hmu_S)
\nn \\
&+& P^{HRG}_{|S|=3,B} \cosh(\hmu_B-3\hmu_S) 
\;,
\label{eq:P-HRG}
\eeqa
within the classical Boltzmann approximation. In the temperature range  $130~{\rm
MeV}\lesssim T\lesssim 200$ MeV relevant for our discussion the Boltzmann
approximation gives at most $3\%$ corrections to the full HRG model results for all
the susceptibilities involving strangeness considered here and hence is well
justified.  Here $\hmu_{B/S}=\mu_{B/S}/T$ are the dimensionless baryon and
strangeness chemical potentials. $P^{HRG}_{|S|=1,M}$ is the partial pressure of all
$|S|=1$ mesons and $P^{HRG}_{|S|=i,B}$ are the partial pressures of all $|S|=i$
($i=1,2,3$) baryons , for $\mu_B=\mu_S=0$. For simplicity, we have set the electric
charge chemical potential $\hmu_Q=0$.

To investigate the sDoF we will use the dimensionless generalized susceptibilities
of the conserved charges
\beq
\chi_{mn}^{XY} = \left. \frac{\partial^{(m+n)} [p(\hmu_X,\hmu_Y)/T^4]} 
{\partial \hmu_X^m \partial \hmu_Y^n} \right|_{\vec{\mu}=0}
\ ,
\label{eq:susc}
\eeq
where $X,Y=B,S,Q$ and $\vec{\mu}=(\mu_B,\mu_S,\mu_Q)$. We also use the notations
$\chi_{0n}^{XY}\equiv\chi_n^Y$ and $\chi_{m0}^{XY}\equiv\chi_m^X$. 

Using the two strangeness fluctuations ($\chi_2^S, \chi_4^S$) and the four
baryon-strangeness correlations ($\chi_{11}^{BS}, \chi_{13}^{BS}, \chi_{22}^{BS},
\chi_{31}^{BS}$) up to fourth order, we have a set of six susceptibilities that can
be used to construct observables that project onto the four different partial
pressures in an uncorrelated hadrons gas introduced in \eq{eq:P-HRG}. 
\beqa
M(c_1,c_2) &=& \chi_2^S -\chi_{22}^{BS} + c_1 v_1 + c_2 v_2 
\;, \label{eq:M} \\
B_1(c_1,c_2) &=&  \frac{1}{2} \left( \chi_4^S - \chi_2^S +5 \chi_{13}^{BS}+
7 \chi_{22}^{BS} \right) \nn \\
&+& c_1 v_1 + c_2 v_2   
\;, \label{eq:B1} \\
B_2(c_1,c_2) &=& - \frac{1}{4} \left( \chi_4^S - \chi_2^S + 4 \chi_{13}^{BS} +
4 \chi_{22}^{BS} \right) \nn \\
&+&  c_1 v_1 + c_2 v_2 
\;, \label{eq:B2}\\
B_3(c_1,c_2) &=& \frac{1}{18} \left( \chi_4^S -  \chi_2^S + 3 \chi_{13}^{BS}+
3 \chi_{22}^{BS} \right) \nn \\
&+& c_1 v_1 + c_2 v_2 
\;. \label{eq:B3}
\eeqa 
The combination $c_1 v_1 + c_2 v_2$ spans a two dimensional plane in the
6-dimensional space of susceptibilities on which the partial pressure $P^{HRG}_S$
vanishes identically when the sDoF are described by a gas of uncorrelated hadrons
irrespective of their masses. The two additional free parameters, $c_1$ and $c_2$,
can thus be used to construct observables that have an identical interpretation in
the uncorrelated hadron gas, but differ under other circumstances, for instance in a
medium where the sDoF are carried by quark-like quasi-particles. For $v_1$ and $v_2$
we choose the following combinations
\beqa
v_1 &=& \chi_{31}^{BS} - \chi_{11}^{BS} 
\;, \label{eq:v1} \\
v_2 &=& \frac{1}{3} (\chi_2^S - \chi_4^S ) - 2 \chi_{13}^{BS} - 
4 \chi_{22}^{BS} - 2 \chi_{31}^{BS} 
\;. \label{eq:v2}
\eeqa
Since in a hadron gas the baryonic sDoF are associated with $|B|=1$, the
baryon-strangeness correlations differing by even numbers of $\mu_B$ derivatives are
identical, giving $v_1=0$. $v_2$ can be re-written as the difference of two operators
\footnote{$3v_2=(\chi_2^S-\chi_{13}^{BS}/6-2\chi_{22}^{BS}-11\chi_{31}^{BS}/6) -
(\chi_4^S-35\chi_{13}^{BS}/6-10\chi_{22}^{BS}-25\chi_{31}^{BS}/6)$} each of which
corresponds to the partial pressure of all strange hadrons in an uncorrelated hadron
gas, leading to $v_2=0$.  Thus, for a classical uncorrelated hadron gas such as the
HRG model $M(c_1,c_2)\to P^{HRG}_{|S|=1,M}$ and $B_i(c_1,c_2)\to P^{HRG}_{|S|=i,B}$
($i=1,2,3$), independent of the values of $c_1$ and $c_2$. For asymptotically high
temperatures, \ie\ when the sDoF are non-interacting massless quarks, these
observables will generically attain different values for different combinations of
$(c_1,c_2)$.


\begin{figure}[!t]
\includegraphics[scale=0.5]{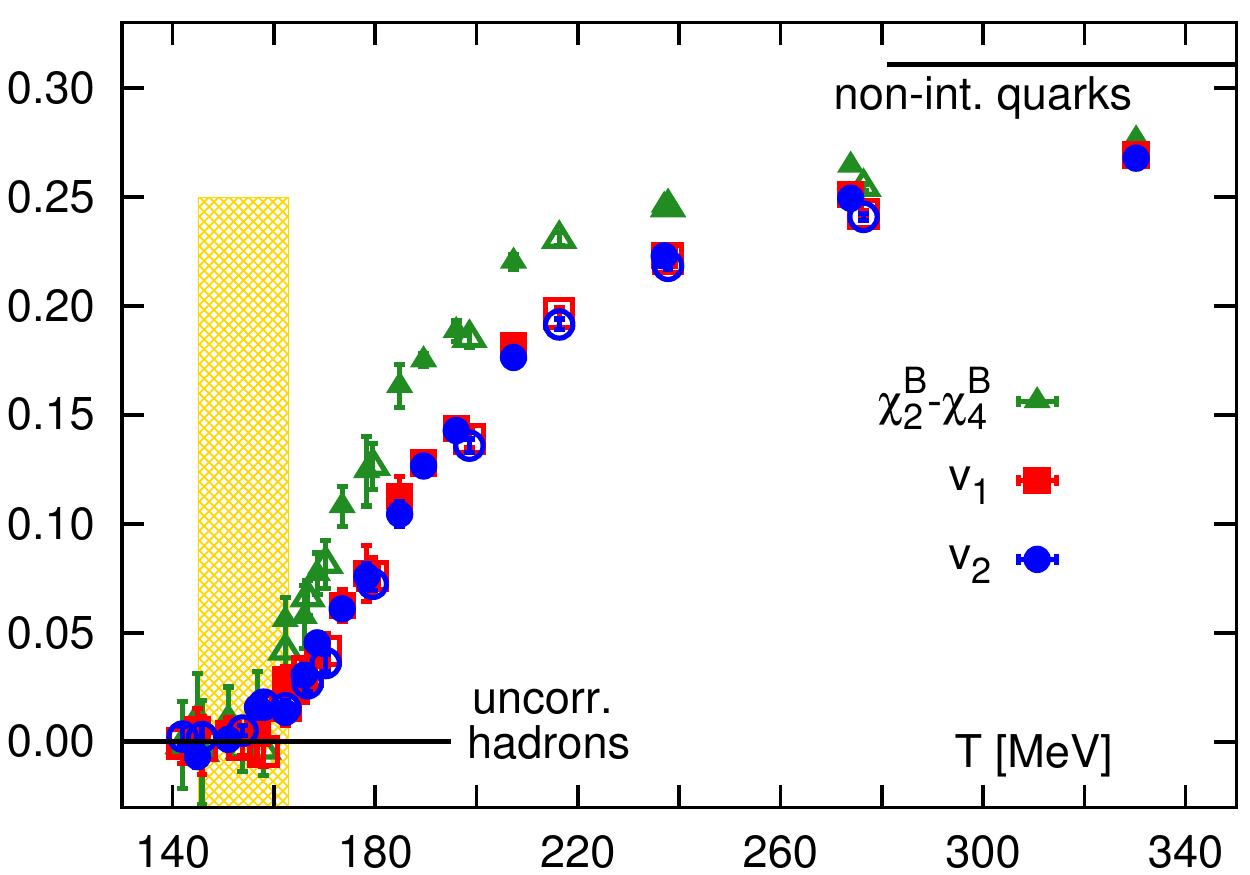}
\caption{Two combinations, $v_1$ and $v_2$ [see Eqs.\ (\ref{eq:v1}-\ref{eq:v2})], of
strangeness fluctuations and baryon-strangeness correlations that vanish identically
if the sDoF are described by an uncorrelated gas of hadrons. Also shown is the
difference of quadratic and quartic baryon number fluctuations, $\chi_2^B-\chi_4^B$.
This observable also vanishes identically when the baryon number carrying degrees of
freedom are described by an uncorrelated gas of strange as well as non-strange
baryons. The shaded region indicates the chiral crossover temperature $T_c=154(9)$
MeV \cite{hotQCDTc}. The lines at low and high temperatures indicate the two limiting
scenarios when the dof are described by an uncorrelated hadron gas and
non-interacting massless quark gas, respectively. The LQCD results for the
$N_\tau=6$ and $8$ lattices are shown by the open and filled symbols respectively.}
\label{fig:v1-v2}
\end{figure}

\begin{figure*}[!t]
\includegraphics[width=0.24\textwidth,height=0.15\textheight]{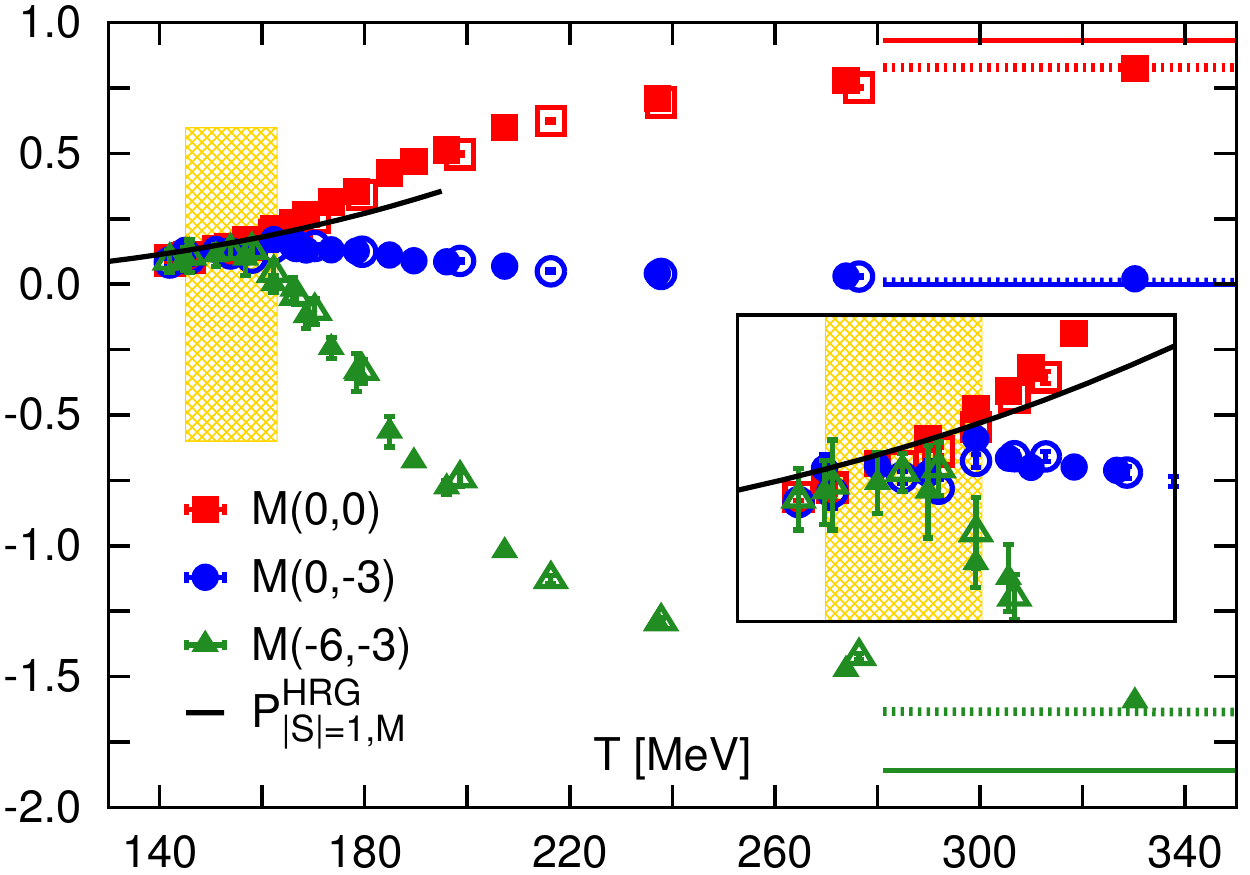}
\includegraphics[width=0.24\textwidth,height=0.15\textheight]{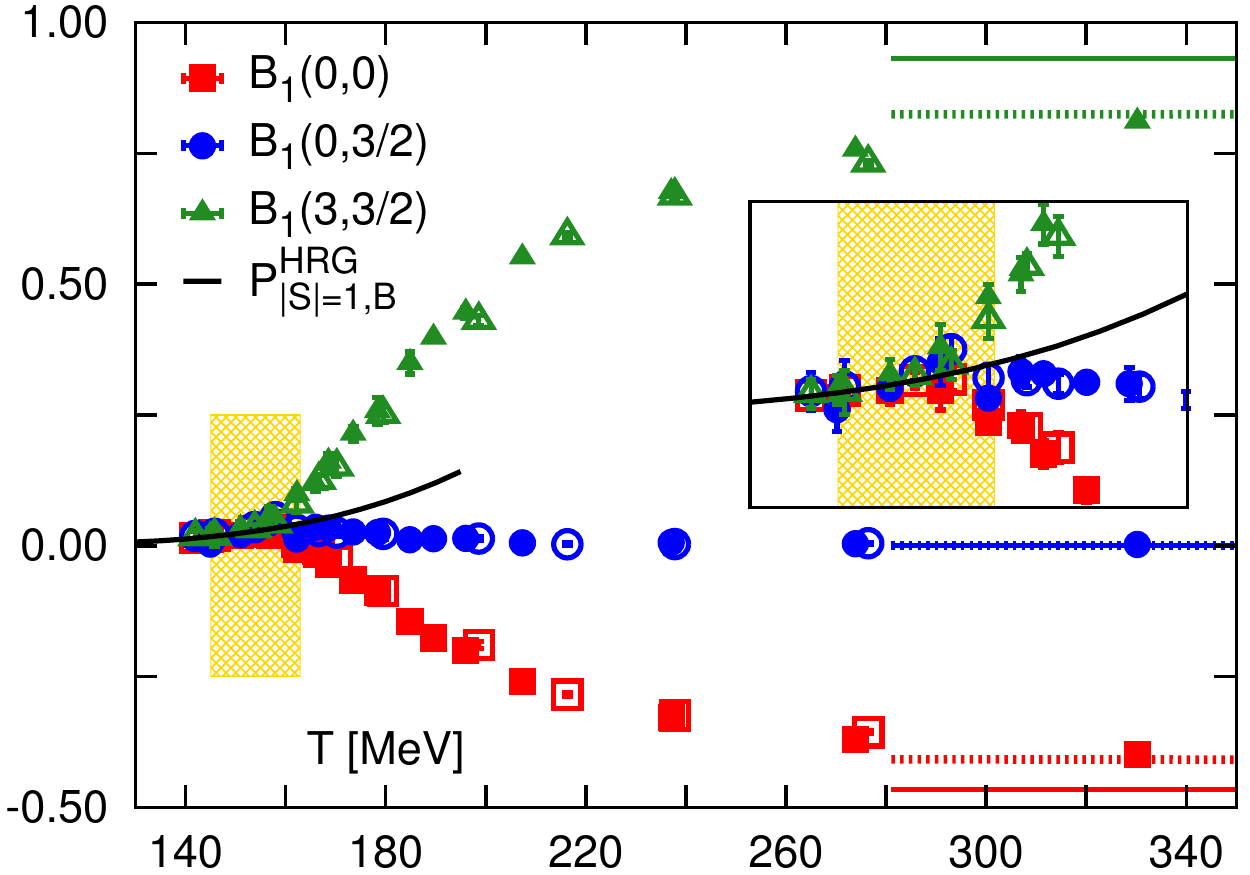}
\includegraphics[width=0.24\textwidth,height=0.15\textheight]{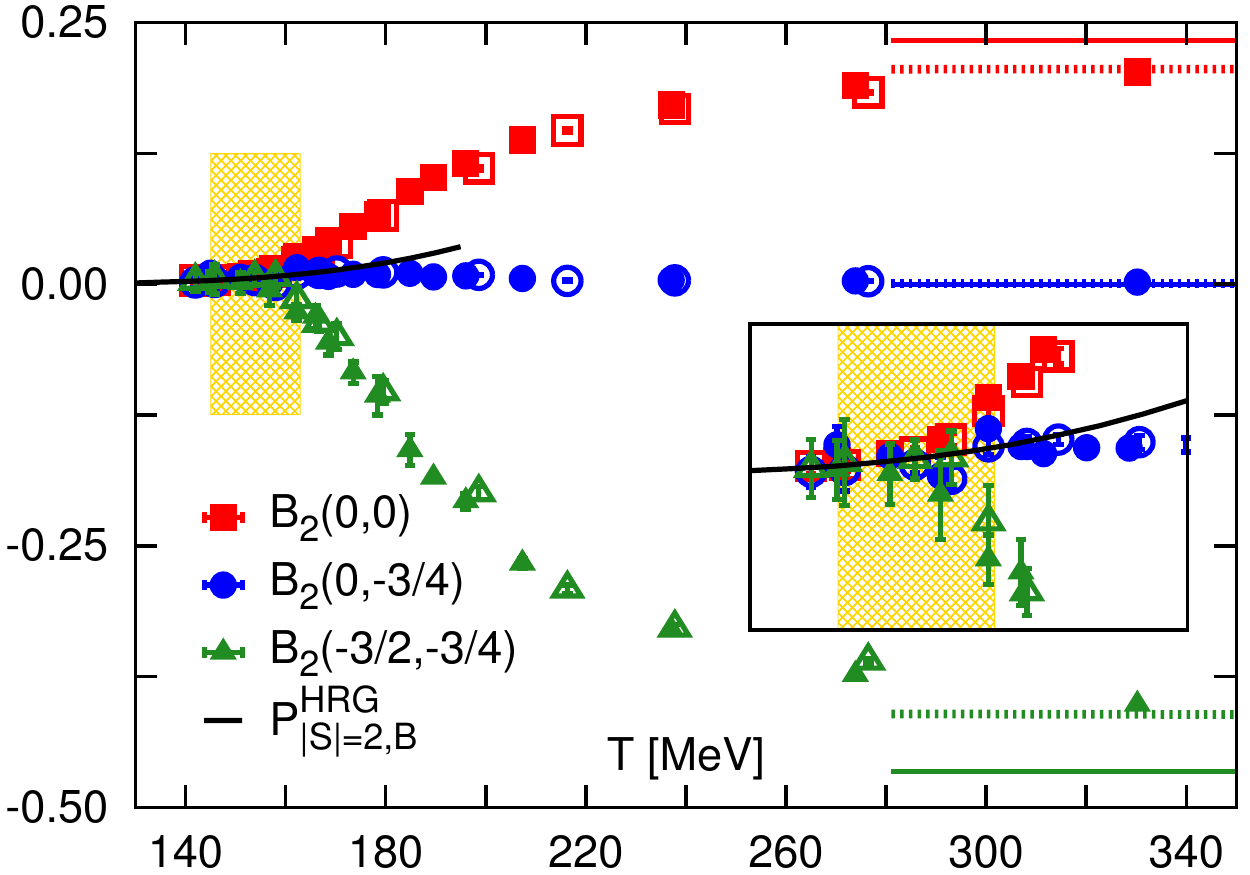}
\includegraphics[width=0.24\textwidth,height=0.15\textheight]{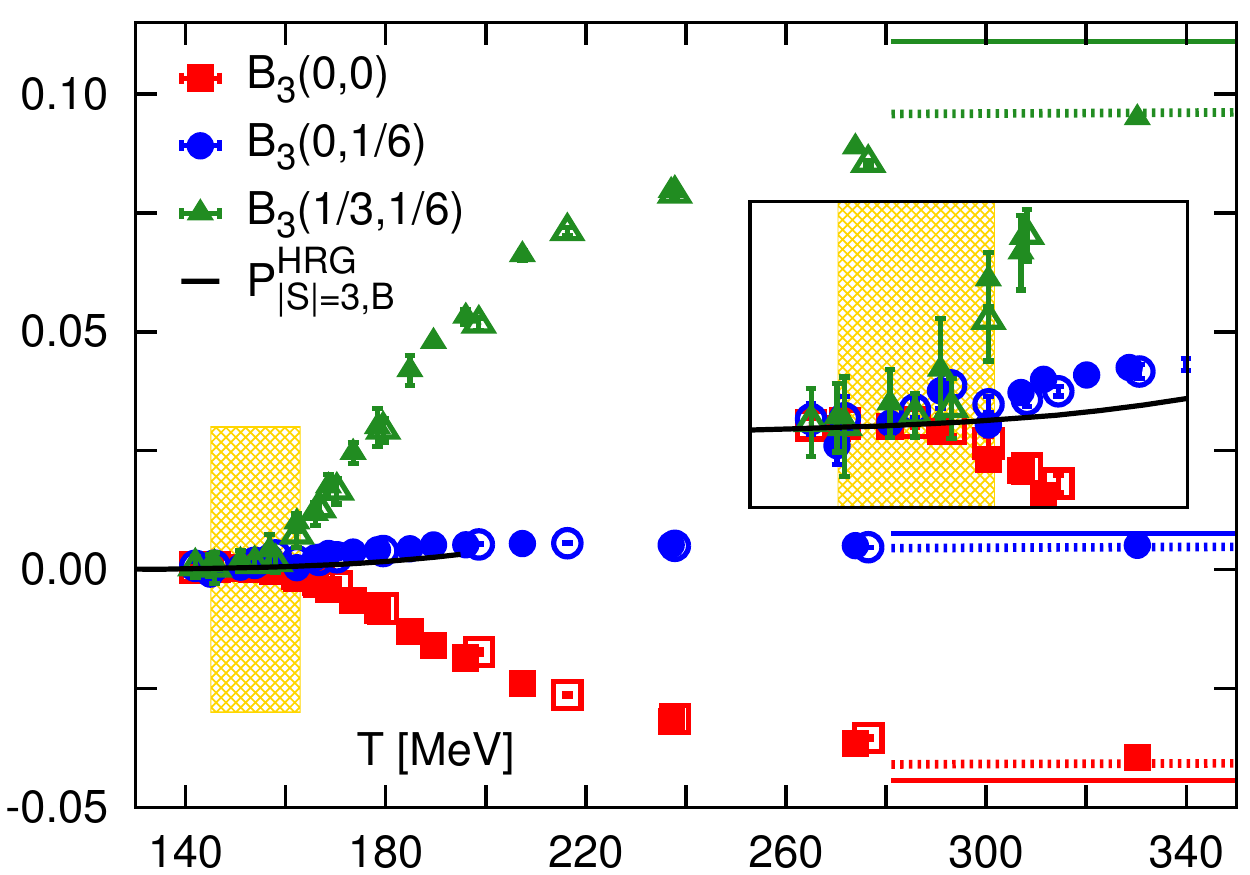}
\caption{Four combinations [see Eqs.\ (\ref{eq:M}-\ref{eq:B3})] of net strangeness
fluctuations and baryon-strangeness correlations $M(c_1,c_2)$, $B_1(c_1,c_2)$,
$B_2(c_1,c_2)$ and $B_3(c_1,c_2)$ (from left to right), each for three different sets
of $(c_1,c_2)$. Up to the chiral crossover temperature $T_c=154(9)$ MeV
\cite{hotQCDTc} (shown by the shaded regions), independent of $(c_1,c_2)$, these
combinations give the partial pressures of $|S|=1$ mesons ($P_{|S|=1,M}^{HRG}$) and
$|S|=1,2,3$ baryons ($P_{|S|=1,B}^{HRG}$, $P_{|S|=2,B}^{HRG}$, $P_{|S|=3,B}^{HRG}$)
in an uncorrelated gas of hadrons having masses equal to their vacuum masses, \ie\ in
the HRG model (indicated by the solid lines at low temperatures). Above the $T_c$
region such a hadronic description breaks down (shown in the insets) and all the
combinations smoothly approach towards their respective, $(c_1,c_2$) dependent, high
temperature limits (indicated by the solid horizontal lines at high temperatures)
described by the non-interacting massless strange quarks. The dotted horizontal lines
at high temperatures depict the perturbative estimates (see text) for all these
observables obtained using one-loop re-summed HTL calculations \cite{Vuorinen}.  The
LQCD results for the $N_\tau=6$ and $8$ lattices are shown by the open and filled
symbols respectively.} 
\label{fig:M-B1-B2-B3} 
\end{figure*}

\emph{Strangeness near the chiral crossover.---} Here we investigate to what extent
sDoF are described by an uncorrelated hadron gas in the vicinity of the chiral
crossover temperature $T_c=154(9)$ MeV \cite{hotQCDTc}. The LQCD results for the
susceptibilities were obtained for two different lattice spacings ($a$) corresponding
to temporal extents $N_\tau=1/aT=6$ and $8$ using $\mathcal{O}(a^2)$ improved gauge
and Highly Improved Staggered Quark \cite{Follana} discretization schemes for $(2+1)$
flavor QCD. The up and down quark masses correspond to a Goldstone pion mass of $160$
MeV and the strange quark mass is tuned to its physical value. The susceptibilities
were measured on $3000-8000$ gauge field configurations, each separated by $10$
molecular dynamics trajectories, using $1500$ Gaussian random source vectors for each
configuration.  Further details of the LQCD computations can be found in
\cite{hotQCDHRG,hotQCDTc}. Although the LQCD results presented here are not obtained
in the limit of zero lattice spacing, the effects of continuum extrapolations are
known to be quite small for our particular lattice discretization scheme, especially
in the strangeness sector \cite{hotQCDHRG}. This will also be substantiated by the
very mild lattice spacing dependence of our results going from the $N_\tau=6$ to the
$N_\tau=8$ lattices. Thus we expect that the continuum extrapolated results will not
alter the physical picture presented in this paper. 

In \fig{fig:v1-v2} we show the LQCD results for the two combinations $v_1$ and $v_2$,
defined in \eq{eq:v1} and \eq{eq:v2}, that vanish identically in an uncorrelated
hadron gas. The LQCD data for these two quantities are consistent with zero up to
$T_c$ and show rapid increase towards their non-interacting massless quark gas values
above the $T_c$ region. In \fig{fig:v1-v2} we also show the difference between the
quadratic ($\chi_2^B$) and the quartic ($\chi_4^B$) baryon number fluctuations that
also receive contributions from the light up and down quarks. In an uncorrelated gas
of baryons the difference $\chi_2^B-\chi_4^B$ also vanishes identically, owing to the
fact that all baryon number carrying degrees of freedom are associated with $|B|=1$.
The LQCD results for this quantity are also consistent with such a hadronic
description up to $T_c$, showing rapid departures above the $T_c$ region. All these
results indicate that till the chiral crossover, the sDoF are in accord with that of
the uncorrelated gas of hadrons and such a description breaks down within the chiral
crossover region. Since $v_1$ is the analog  of $\chi_2^B-\chi_4^B$ in the strange
baryon sector, the fact that both quantities have similar temperature dependence
shows that the behavior of the sDoF around the chiral crossover region is rather akin
to the behavior of degrees of freedom involving the light quarks. Note that the
vanishing values of these observables at low temperatures are independent of the mass
spectrum of the hadrons, as long as they are uncorrelated and the Boltzmann
approximation is well suited. It reflects that the relevant degrees of freedom carry
integer strangeness $|S|=0,1,2,3$ and integer baryon number $|B|=0,1$.

In \fig{fig:M-B1-B2-B3}, we study the partial pressures of the strange hadrons using
the LQCD results for the four combinations $M(c_1,c_2)$, $B_1(c_1,c_2)$,
$B_2(c_1,c_2)$ and $B_3(c_1,c_2)$ (see Eqs.\ (\ref{eq:M}-\ref{eq:B3})), each for
three sets of $(c_1,c_2)$. One of the combinations corresponds to $c_1=c_2=0$ and
thus represents the basic projection onto a given strangeness sector in an
uncorrelated hadron gas. The other two parameter sets for $(c_1,c_2)$ are chosen to
produce widely different values for these observables in a non-interacting massless
quark gas at asymptotically high temperatures. From \fig{fig:v1-v2} it is obvious
that they are identical at low temperatures. Up to $T_c$, independent of $(c_1,c_2)$,
these four quantities individually agree with the partial pressures of the $|S|=1$
mesons and the $|S|=1,2,3$ baryons when one uses the actual vacuum mass spectrum of
the strange hadrons in an uncorrelated hadron gas. Specifically, $M(c_1,c_2)$,
$B_1(c_1,c_2)$, $B_2(c_1,c_2)$ and $B_3(c_1,c_2)$ reproduce the HRG model results for
$P_{|S|=1,M}^{HRG}$, $P_{|S|=1,B}^{HRG}$, $P_{|S|=2,B}^{HRG}$ and $P_{|S|=3,B}^{HRG}$
respectively \footnote{In the HRG model calculations we have used all the three star
hadrons with masses $\leq2.5$ GeV as listed in the 2010 summary table of the Particle
Data Group (PDG) \cite{Nakamura}. We have checked the HRG results by taking into
account higher mass hadrons and by reducing the mass cut-off to $2$ GeV. Such changes
give results which are at most a couple of percent different in the relevant
temperature range and do not alter our conclusions.}. As can be seen from the insets
of \fig{fig:M-B1-B2-B3}, such a description of the LQCD results breaks down within
the $T_c$ region for each of the meson and baryon sector. Above $T_c$, all these
quantities show a smooth approach towards their respective non-interacting, massless
quark gas values depending on the values of $c_1$ and $c_2$.


\emph{Strangeness in the quark gluon plasma.---} To investigate whether the sDoF in
the QGP can be described by weakly interacting quasi-quarks we study correlations of
net strangeness fluctuations with fluctuations of net baryon number and electric
charge. Such observables were studied in \cite{Koch,Majumder:2006nq} for the second
order correlations. We extend these correlations up to the fourth order. If the sDoF
are weakly/non-interacting quasi-quarks then strangeness $S=-1$ is associated with
the fractional baryon number $B=1/3$ and electric charge $Q=-1/3$ giving
\beq
\frac{\chi_{mn}^{BS}}{\chi_{m+n}^S}=\frac{(-1)^n}{3^m}
\;,\qquad\mathrm{and}\qquad
\frac{\chi_{mn}^{QS}}{\chi_{m+n}^S}=\frac{(-1)^{m+n}}{3^m}
\;,
\label{eq:BS-QS}
\eeq
where $m,n>0$ and $m+n=2,4$. 

In \fig{fig:BS-QS}, we show the LQCD results for these ratios scaled by the proper
powers of fractional baryonic and electric charges. The shaded regions at high
temperatures indicate the ranges of values for these ratios as predicted for the
weakly interacting quasi-quarks from the re-summed Hard Thermal Loop (HTL)
perturbation theory at the one-loop order \cite{Vuorinen}, using one-loop running
coupling obtained at the scales between $\pi T$ and $4\pi T$. The ratios of the
second order correlations $\chi_{11}^{BS}/\chi_2^S$ and $\chi_{11}^{QS}/\chi_2^S$ are
much closer to those expected for weakly interacting quasi-quarks, differing only at
a few percent level for $T\sim1.25T_c$. Previous LQCD studies
\cite{BSold,WHRG,hotQCDHRG} also showed similar results, suggesting that sDoF in the
QGP can be described by weakly interacting quasi-quarks even down to temperatures
very close to $T_c$. However, our results involving correlations of strangeness with
higher power of baryon number and electric charge clearly indicate that such a
description in terms of weakly interacting quasi-quarks can only be valid for
temperatures $T\gtrsim2T_c$. While the HTL perturbative expansion for ratios
involving one derivative of the baryonic/electric charges (\ie\
$\chi^{XS}_{11}/\chi_2^S$ and $\chi^{XS}_{13}/\chi_4^S$, $X=B,Q$) starts differing
from the non-interacting quark gas limit at $\mathcal{O}(\alpha_s^3\ln\alpha_s)$
\cite{Blaizot}, the same for those involving higher derivatives of the
baryonic/electric charges (\ie\ $\chi^{XS}_{22}/\chi_4^S$ and
$\chi^{XS}_{31}/\chi_4^S$, $X=B,Q$) starts at $\mathcal{O}(\alpha_s^{3/2})$
\cite{Vuorinen}, $\alpha_s$ being the strong coupling constant. Thus, the enhancement
of the higher order electric charge/baryon-strangeness correlations is probably
expected within the regime of validity of the weak coupling expansion. Apart from the
charge/baryon-strangeness correlations presented here, at the fourth order there are
$3$ more susceptibilities involving derivatives with respect to all $B$, $Q$ and $S$,
namely $\chi_{ijk}^{BQS}$ with $i,j,k>0$ and $i+j+k=4$. These
baryon-charge-strangeness correlations can also be used in a similar manner to
simultaneously probe both $B$ and $Q$ associated with the sDoF. We have studied these
quantities too and arrived at the same conclusion regarding the validity of the
weakly interacting quark gas picture at high temperatures. Also note that the high
temperature behavior of the quantities depicted in \fig{fig:M-B1-B2-B3} reaffirms our
conclusion that the weakly interacting quark gas picture can only be valid for
$T\gtrsim2T_c$. For temperatures beyond the validity of the weak coupling expansion,
it would be interesting to see whether such enhancements indicate a strongly coupled
QGP \cite{str-coup} without quasi-particles or signal the presence of colored bound
states \cite{Shuryak} and/or density dependent massive quasi-particles
\cite{quasipart2}.

\begin{figure}[!t]
\includegraphics[scale=0.5]{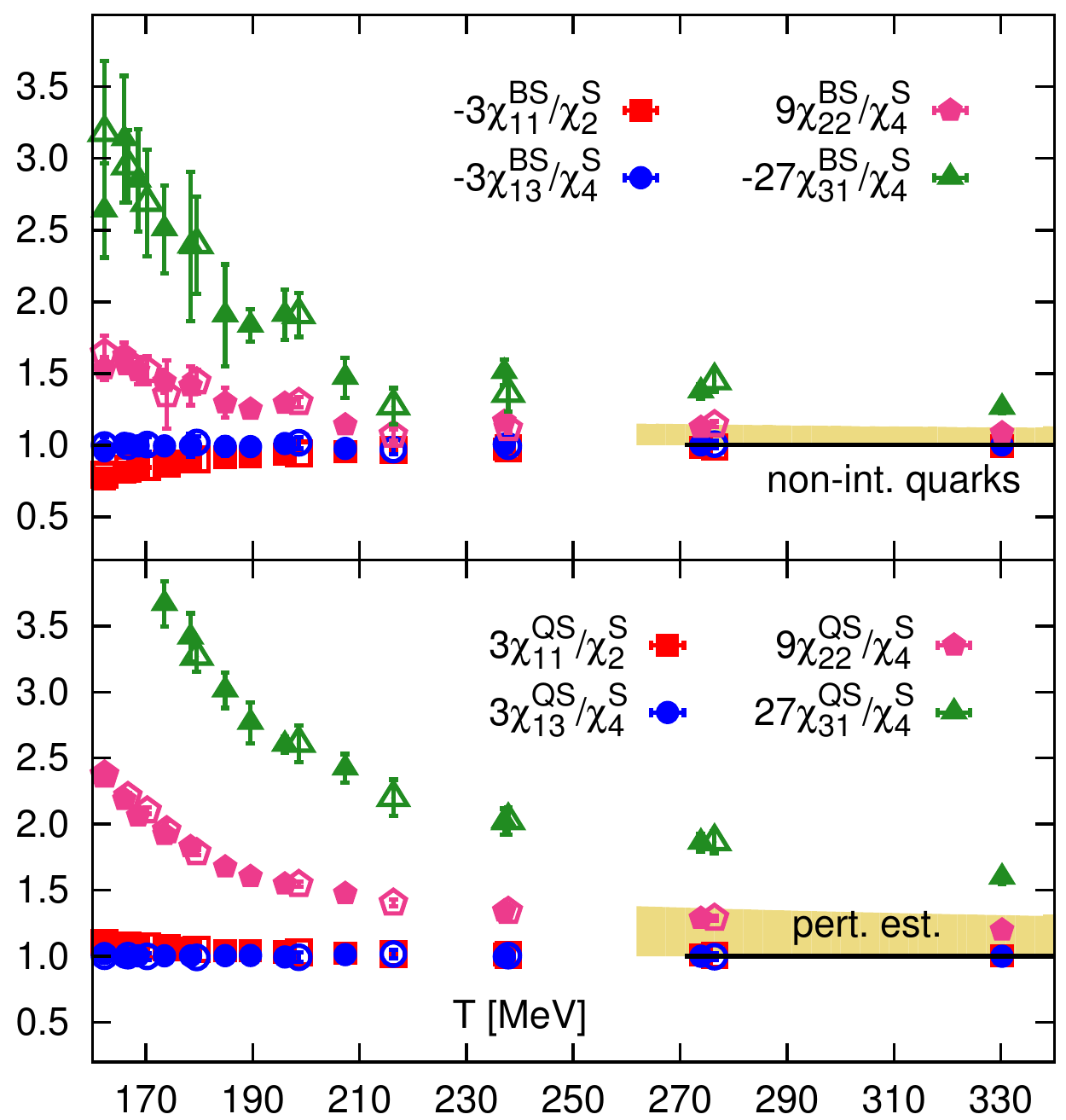}
\caption{Baryon-strangeness (top) and electric charge-strangeness correlations
(bottom), properly scaled by the strangeness fluctuations and powers of the fractional
baryonic and electric charges [see \eq{eq:BS-QS}]. In the non-interacting massless
quark gas all these observables are unity (shown by the lines at high temperatures).
The shaded regions indicate the range of perturbative estimates (see text) for all
these observables obtained using one-loop re-summed HTL calculations \cite{Vuorinen}.
The LQCD results for the $N_\tau=6$ and $8$ lattices are shown by the open and filled
symbols respectively.}
\label{fig:BS-QS}
\end{figure}


\emph{Conclusions.---} The LQCD results presented in this paper show that till the
chiral crossover temperature $T_c$ the quantum numbers associated with the sDoF are
consistent with those of an uncorrelated gas of hadrons. Furthermore, up to $T_c$ the
partial pressures of the strange mesons and baryons are separately in agreement with
those obtained from the uncorrelated hadron gas using vacuum masses of the strange
hadrons. Such a hadronic description of the sDoF breaks down in the chiral crossover
region. Moreover, the behavior of the sDoF around $T_c$ is quite similar to that
involving the light up and down quarks. Altogether, these results suggest that the
deconfinement of strangeness seemingly takes place at the chiral crossover
temperature. On the other hand, our LQCD results involving correlations of
strangeness with higher powers of baryonic and electric charges for $T>T_c$ provide
unambiguous evidence that the sDoF in the QGP can become compatible with the weakly
interacting quark gas only for $T\gtrsim2T_c$. For the intermediate temperatures,
$T_c\lesssim T\lesssim2T_c$, strangeness is non-trivially correlated with the
baryonic and electric charges indicating that the QGP in this temperature regime
remains strongly interacting.


\emph{Acknowledgments.---} This work has been supported in part through contract
DE-AC02-98CH10886 with the U.S. Department of Energy, through Scientific Discovery
through Advanced Computing (SciDAC) program funded by U.S. Department of Energy,
Office of Science, Advanced Scientific Computing Research and Nuclear Physics, the
BMBF under grant 05P12PBCTA, the DFG under grant GRK 881, EU under grant 283286 and
the GSI BILAER grant. Numerical calculations have been performed using the USQCD
GPU-clusters at JLab, the Bielefeld GPU cluster and the NYBlue at the NYCCS. We also
thank Nvidia for supporting the code development for the Bielefeld GPU cluster.


\end{document}